\documentclass[aps,twocolumn,nofootinbib]{revtex4-2}

\usepackage{graphicx}
\usepackage{dcolumn}
\usepackage{bm}
\usepackage{array}
\usepackage{booktabs}
\usepackage{multirow}
\newcommand{\head}[1]{\textnormal{\textbf{#1}}}
\newcommand{\normal}[1]{\multicolumn{1}{l}{#1}}
\usepackage{amssymb,amsfonts}
\usepackage{epsfig}
\usepackage{epstopdf}
\usepackage[all]{xy}
\usepackage{amsthm}
\usepackage{dcolumn}
\usepackage{hyperref}
\usepackage{url}
\usepackage{dsfont}
\usepackage{slashed}
\usepackage{mathrsfs}

\newcommand{\be}{\begin{equation}}
\newcommand{\ee}{\end{equation}}
\newcommand{\bea}{\begin{eqnarray}}

\newcommand{\eea}{\end{eqnarray}}

\def\inbar{\,\vrule height1.5ex width.4pt depth0pt}
\def\IR{\relax{\rm I\kern-.18em R}}
\def\IC{\relax\hbox{$\inbar\kern-.3em{\rm C}$}}

\begin{document}

\title{ Production cross section of heavy quarks in $e^{-}p$ interaction at the NLO approximation}

\author{S. Zarrin\footnote{zarrin@phys.usb.ac.ir}}

\author{S. Dadfar}

\author{M. Sayahi}

\affiliation{ Department of Physics, University of Sistan and Baluchestan, Zahedan, Iran}

\begin{abstract}
We present the production cross section of heavy quarks $\sigma^{c\bar{c}}$, $\sigma^{b\bar{b}}$ and $\sigma^{t\bar{t}}$ at the next-to-leading order in the electron-proton interaction by using the  quarks and gluon distribution functions at the initial scale $Q_{0}^{2}$. To do this, we use a fitted form of the heavy quark coefficient functions for
deep-inelastic lepton-hadron scattering to obtain the structure functions of heavy quarks. Then, we calculate the reduced cross section of heavy quarks by using the structure functions and subsequently present the single differential and the integrated cross section of heavy quarks  at the center-of-mass energies of 
$\sqrt{s}=319$ $GeV$, $1.3$ $TeV$ and  $3.5$ $TeV$ in the electron-proton collision. The obtained numerical results of the cross section of the charm and beauty quarks are compared with the HERA data, which is a combination from the results of the H1 and ZEUS detectors, and with the predictions from  H1PDF, MSTW2008 and MSRT03. Furthermore, we  present the production cross section of top quark as a direct prediction from our calculations.
\newline

PACS numbers: 13.60.Hb, 13.85.Lg, 14.65.Dw, 14.65.Fy, 14.65.Ha
 
\end{abstract}
\maketitle
\section{Introduction}
The study of the heavy quarks production is one of the most important subjects of research at the present and future colliders and the test of quantum chromodynamics (QCD). These quarks can be generated in the hadron-hadron, photon-hadron, electron-positron and lepton-hadron interactions. The production of heavy quarks are studied in two different prescriptions in the framework of QCD analyses. The first framework (the region $Q^{2}= m^{2}_{q}$) is the so-called variable-flavour number scheme (VFNS) \cite{1}. In this scheme, the heavy quarks contributions are described by a parton density and treated as a massless quark in the hadron. In the ‘massless’ scheme, the dominant contribution at the leading order (LO) approximation is due to the quark parton model (QPM) process and at the next-to-leading order (NLO) approximation the contributions of the photon-gluon fusion (PGF) and QCD Compton processes are also considered.   In the second framework,  heavy quarks are treated as a massive quark and their contributions are given by fxed-order perturbation theory (FOPT)\cite{2,3}.  In this scheme (the ‘massive’ scheme), the dominant LO process is the PGF and the NLO diagrams are of order $\alpha_{s}^{2}$. 

At HERA, at the LO approximation, the PGF is the dominant contribution for the heavy quarks production in electron-proton interaction $e^{-}+p\rightarrow q\bar{q}+e^{-}+X$. In this process, by the interaction of a virtual photon emitted by the incoming electron with a gluon from the proton,  is formed a heavy quark-antiquark pair. The HERA data show that the production of heavy quarks is sensitive to the gluon distribution (which the minimum momentum fraction of gluon $x_{g}$ in photoproduction to generate a heavy quark pair is  arranged  such  that $x^{tt}_{g} > x^{bb}_{g }> x^{cc}_ {g}$)  and also is dependent on the mass  of these quarks. Therefore,  the  calculations of the heavy quarks structure functions are dependent on the squared energy scale $\mu^{2}$ \cite{8,9,10,11,12}. 

The measurements of the open charm (c) cross section in DIS at HERA have mainly been exclusive for $D$ or $D^{*}$ meson production \cite{15,151,152,153,154}. The measurement of the open beauty (b) cross section is challenging since b events contain only a small fraction (typically $< 5\%$) of the total cross section. The $b$ cross section has been measured in DIS ($Q^2 > 2GeV^2$) by ZEUS \cite{17} and in photoproduction ($Q^2<1GeV^{2}$ and $0.1<y<0.8$) by H1 \cite {18} and ZEUS \cite{19}, using the transverse momentum distribution of muons relative to the $b$ jet in semi-muonic decays. Moreover, in Ref. \cite{9}, the production of c and b quarks in $ep$ interactions has been studied with the ZEUS detector at HERA for exchanged four-momentum squared $5 <Q^{2} < 1000GeV^{2}$ using an integrated luminosity of $354 pb^{-1}$. Also, measurements of the c and b contributions to the inclusive proton structure function $F_{2}$ have been recently presented in deep inelastic scattering (DIS) at HERA, using information from the H1 vertex detector, for values of the negative square of the four momentum of the exchanged boson $Q^2> 150 GeV^2$  and of inelasticity $0.1 <y< 0.7$ \cite{13}. In this region, the inclusive c and b cross sections have been found  $\sigma_{c\bar{c}}= 373 \pm 39 \pm 47$ $pb$ and  $\sigma_{b\bar{b}} =55.4\pm8.7\pm12.0$ $pb$, respectively and the data show that a fraction of $\sim 18\%$ ($\sim 3\%$) of DIS events contain c (b) quark. Furthermore, inclusive c and b cross sections have been measured in $e^{-}p$ and $e^{+}p$ neutral current collisions at HERA in the kinematic region of  $5<Q^{2}<2000 GeV^{2}$ and $0.0002<x< 0.05$ which $x$ is the Bjorken scaling variable. In which the $e^{-}p$ center-of-mass energy (CME) is $\sqrt{s} = 319 GeV$, with a proton beam energy of $E_{p} =920 GeV$ and electron beam energy of $E_{e} = 27.6 GeV$ \cite{8}.

In high energy processes, the contribution of heavy quarks in the proton structure functions will be studied in projects such as the Large Hadron electron Collider (LHeC) and the Future Circular Collider electron-hadron (FCC-eh) which operate at high enough energies to observe new phenomenon \cite{22,23,24,25,26,27}. At the LHeC project, the possibility of colliding an electron beam from a new accelerator with the existing LHC proton is investigated.  In this project,  the  $e^{-}p$  CME is planned to reach $\sqrt{s}=1.3TeV$  \cite{22,26}. Beyond the LHeC, the next generation $ep$ collider (the FCC-eh project) is an ideal environment to increase center-of-mass energy.  In the proposed FCC-eh program, the distribution of heavy quarks  will be examined at $\sqrt{s}=3.5TeV$ \cite{27}. 
 
Theoretically, the inclusive heavy quark production have been presented within the VFNS scheme at the next-to-next-to-leading order approximation (NNLO) in Ref. \cite{4}. The predictions for the c and b cross sections have been obtained from fits \cite{5} to the HERA inclusive $F_{2}$ data based on CCFM evolution \cite{6}. Also, the production of heavy quarks in the FFNS approach have been predicted  according to the LO PGF off-shell matrix elements convoluted with the CCFM $k_{T}$-unintegrated gluon density of the proton \cite{5}. In Refs. \cite{28,29,30,31,32,321}, the connection between the gluon distribution and the structure  functions of heavy quarks (c and b) has been theoretically shown at small $x$. Moreover, in Refs. \cite{33,34,321} the authors present the  necessary conditions for predicting the top structure function $F^{t\bar{t}}_{2}$ with respect to the different predictions for the behavior of the gluon at low $x$ and high  $Q^{2}$ values. Besides these predictions, various successful phenomenological methods have presented to obtain the c and b structure functions and the ratios of $R^{c\bar{c}}$ and $R^{b\bar{b}}$ \cite{29,35,321}. The importance of these studies,  along  with  the  $t$-quark density, can  be  explored  at  future circular collider energies  and may lead us to new physics in the future \cite{36,37}.

At small $x$, both the H1 and ZEUS detectors have measured the charm and beauty components of the proton structure function from the measurement of the inclusive heavy quark cross sections after applying small corrections to the longitudinal structure function of heavy quarks at low and moderate inelasticity. But, in the region of high inelasticity, this function may has a significant effect on the heavy quark production cross section.  The heavy quark deferential cross section is written in terms of the heavy quark structure functions as:

$$
\frac{d^{2}\sigma^{q\bar{q}}}{dxdQ^{2}}=\frac{2\pi\alpha^{2}}{xQ^{2}}\bigg(Y_{+}F_{2}^{q\bar{q}}(x,Q^{2})-y^2F_{L}^{q\bar{q}}(x,Q^{2})\bigg)\nonumber
$$
\begin{equation}
=\frac{2\pi\alpha^{2}}{xQ^{2}}Y_{+}\sigma^{q\bar{q}}_{red}(x,Q^{2}),
\end{equation}
where $Y_{+}=1+(1-y)^{2}$ and $y=Q^{2}/(xs)$ is the inelasticity variable in which $s$  and $Q^{2}$ are the CME squared and the photon virtuality, respectively. The heavy quark structure functions $F_{2}^{q\bar{q}}(x,Q^{2})$ and $F_{L}^{q\bar{q}}(x,Q^{2})$ with respect to the behavior of the gluon density are given by:
\begin{equation}
F_{k,g}^{q\bar{q}}(x,Q^{2},m_{q}^{2})=xe_{H}^{2}\int_{x}^{z_{max}}H_{k,g}(z,\xi) g(\frac{x}{z},\mu^{2})\frac{dz}{z},
\end{equation}
where $\mu=(Q^2+4m_{q}^{2})^{1/2}$ is the default common value for the factorization and renormalization scales, $z_{max}=\frac{Q^{2}}{\mu^{2}}$ and $\xi=\frac{Q^{2}}{m_{q}^2}$. In general, the heavy quark coefficient functions of $H_{k,g}(z,\xi)$ (with $k=2,L$) are expanded in $\alpha_{s}$ as follows:
\begin{equation}
H_{k,g}(z,\xi)=\sum_{i=1}^{\infty}\left(\frac{\alpha_{s}(\mu^{2})}{4\pi}\right)^{i}H^{(i)}_{k,g}(z,\xi), \ \ k=2,L,
\end{equation}
where the heavy quark coefficient functions at the LO and NLO approximations, $H_{k,g}^{(1)}$ and $H_{k,g}^{(2)}$, are as follows:
$$
H_{k,g}^{(1)}(z,\xi)=\frac{\xi}{\pi z}c^{(0)}_{k,g}(\eta,\xi),
$$
\begin{equation}
H_{k,g}^{(2)}(z,\xi)=\frac{16 \pi \xi}{ z}\left[c^{(1)}_{k,g}(\eta,\xi)+\bar{c}^{(1)}_{k,g}(\eta,\xi)\ln\left(\frac{\mu^{2}}{m_{q}^{2}}\right)\right],
\end{equation}
where the coefficient functions $c^{(0)}_{k,g}(\eta,\xi)$ have been given in Ref. \cite{3} and  the coefficients $c^{(1)}_{k,g}$ and $\bar{c}^{(1)}_{k,g}$ are rather lengthy, and not published in print and they are only available as computer codes \cite{3}. In Ref. \cite{38}, the analytic form of the heavy quark coefficient functions have been presented  for deep-inelastic lepton-hadron scattering in the kinematical regime $Q^{2}\gg m^{2}_{q}$ in which $Q^{2}$ and $m^{2}_{q}$ stand for the masses squared of the virtual photon and heavy quark, respectively.

In  this paper, we obtain heavy quark structure functions $F_{2}^{q\bar{q}}$ and  $F_{L}^{q\bar{q}}$ at the LO and NLO approximations. These functions are obtained by using the presented heavy quark coefficient functions in Ref. \cite{38}  (in the kinematical regime $Q^{2}\gg m^{2}_{q}$) and a set control coefficients which are obtained by using the heavy quark structure functions from  HERA \cite{9,13,18,39,391,42,421,422}, LHeC \cite{43}, other works such as Ref. \cite{44} (for c and b quarks) and Ref. \cite{32} (for t quark) and also the CT18 distribution functions \cite{63}. Then, we present  the integrated  and differential cross sections for heavy quarks in DIS and compare our numerical results with HERA data \cite{8,9,13,17} and with the results from the MSTW2008 \cite{48}, MSRT03 \cite{49}  and H1PDF  \cite{61}.

This paper is organized as follows. In the next section, we present a brief summary of our previous work, including the structure functions of heavy quarks. In section III, we present a detailed numerical analysis of cross sections of heavy quarks. In the last section, we summarize our main conclusions and remarks.

\section{Theoretical formalism}
The presented heavy quark coefficient functions in Ref. \cite{38} at the regimes of $Q^2\geq m^{2}_{q}$ and  $Q^2\leq m^{2}_{q}$ do not provide appropriate and acceptable results and the  structure functions obtained by using these coefficient functions are very ascendant at the regimes of $Q^2\geq m^{2}_{q}$ and  $Q^2\leq m^{2}_{q}$, especially for the charm quark. To control this growth, by using HERA data \cite{9,13,18,39,391,42,421,422} and the results from  Refs. \cite{43,44} (for c and b quarks) and Refs. \cite{32} (for t quarks) and the published CT18 initial distribution functions \cite{63}, we present a series of control coefficients that are only a function of $Q^{2}$. By multiplying these coefficients by the obtained coefficient functions in Ref. \cite{38}, they have given acceptable results. The general form of these control coefficients is as $d-\exp(eQ^{2})$, which $d$  and $e$ are  fixed numbers and are shown in Table (1) for heavy quarks.

 \begin{table}[!]
\begingroup
\fontsize{10pt}{12pt}\selectfont{
\label*{ Table (1): The fixed numbers $d$ and $e$  in the control coefficients of the heavy quark coefficient functions. }}
\newline
\endgroup
\begingroup
\fontsize{8pt}{12pt}\selectfont{
\begin{tabular} {ccc}
\toprule[1pt]
\quad\quad\quad\quad& \normal{\quad \quad\head{$d $}}&
\normal{\quad \quad\quad \quad\head{$e $}}\  \\ \hline
$m_{c}$\quad \quad & $0.975253$\quad\quad&  $-11.9235\times 10^{-4}$   \\ \\
$m_{b}$ \quad\quad& $1.003669$\quad\quad &  $-2.35842\times 10^{-4}$   \\ \\  
$m_{t}$\quad\quad & $1.000011$ \quad\quad&  $-8.93697\times 10^{-8}$   \\  
    
\bottomrule[1pt]
\end{tabular} }
\endgroup
\end{table}

To solve the Eq. (2), it is necessary to obtain the gluon density function, for this aim, we use the DGLAP evolution equations and the Laplace transform method as Ref. \cite{54,55,56} by considering this fact that the Laplace transform of the convolution factors is simply the ordinary product of the Laplace transform of the factors. The coupled DGLAP integral-differential equations are as follows \cite{57}:
$$
 \frac{\partial F_{s}(x,Q^{2})}{\partial \ln Q^{2}}=\frac{\alpha_{s}(Q^{2})}{2\pi}\bigg[P_{qq}(x,Q^{2})\otimes F_{s}(x,Q^{2})
 $$
\begin{equation}
 +2n_{f}P_{qg}(x,Q^{2})\otimes G(x,Q^{2})\bigg],\
\end{equation}
$$
\frac{\partial G(x,Q^{2})}{\partial \ln Q^{2}}=\frac{\alpha_{s}(Q^{2})}{2\pi}\bigg[P_{gq}(x,Q^{2})\otimes F_{s}(x,Q^{2})
$$
\begin{equation}
+P_{gg}(x,Q^{2})\otimes G(x,Q^{2})\bigg], \quad
\end{equation}
where $\alpha_{s}$ is the running strong coupling constant and $P_{ab}(x)$'s  are the Altarelli-Parisi splitting functions. In the above equations, the symbol $\otimes$ represents the convolution integral which is defined as $f(x)\otimes h(x)=\int_{x}^{1}f(y)h(x/y)dy/y$. To convert Eqs. (5) and (6) into Laplace space,  we insert the variables $x=\exp(-v)$, $y=\exp(-w)$ and $\tau(Q^{2},Q_{0}^{2})=\frac{1}{4\pi}\int_{Q^{2}_{0}}^{ Q^{2}} \alpha_{s}(Q{'}^{2})d\ln(Q{'}^{2})$ into the DGLAP evolution equations. By using the Laplace transform method, one can turn the convolution equations at the LO and NLO approximations from $v$-space into $r$-space, and then solve them straightforwardly in $r$-space as:
\begin{equation}\label{eq:11}
f^{(i)}(r,Q^{2})=k_{ff}^{(i)}(r,\tau)f^{(i)}(r,Q_{0}^{2})+k_{fg}^{(i)}(r,\tau)g^{(i)}(r,Q_{0}^{2}),
\end{equation}
\begin{equation}\label{eq:12}
g^{(i)}(r,Q^{2})=k_{gf}^{(i)}(r,\tau)f^{(i)}(r,Q_{0}^{2})+k_{gg}^{(i)}(r,\tau)g^{(i)}(r,Q_{0}^{2}),
\end{equation}
with $i=$ LO or NLO. The functions of $f(r,Q_{0}^{2})$ and $g(r,Q_{0}^{2})$ are the singlet and gluon distribution functions at initial scale $\tau=0$ and $\mathcal{L}[\hat{H}(v,\tau),v,r]= h(r,\tau)$. In Eqs. (\ref{eq:11}) and (\ref{eq:12}), the kernels of $k_{ij}(r,u)$'s at LO and NLO approximations can be found in Refs. \cite{56,561,562}. Since the obtained gluon distribution function in the above equation (8) is in Laplace space $s$ and its exact solution is not possible through analytical techniques, so its inverse Laplace transform must be computed numerically \cite{561,58}.  To obtain the heavy quarks structure functions in terms of the distribution functions at the initial scale, we turn Eq. (2) to the Laplace space $r$. To this aim, the variable $z=x/y$ and  transformation $x\rightarrow xe^{(\ln 1/a)}$ (where $a$ is larger than one) are used, therefore, Eq. (2) is obtained as follows:

$$
F_{k,g}^{q\bar{q}}(xe^{(\ln 1/ a)},Q^{2},m_{q}^{2})=e_{H}^{2}\int_{x}^{1}G(y,\mu^{2})\frac{dy}{y}
$$
\begin{equation}
\times C_{k,g}(\frac{xe^{(\ln 1/ a)}}{y},\xi), \quad k=2, L,
\end{equation}
where  $G(y,Q^{2})=yg(y,Q^{2})$ and $C_{k,g}(x)=xH_{k,g}(x)$. By using the variables $x=\exp(-v)$, $y=\exp(-w)$, one can rewrite the above equation as: 

$$
\hat{F}_{k,g}^{q\bar{q}}(v-\ln 1/a,Q^{2},m_{q}^{2})=e_{H}^{2}\int_{0}^{v}\hat{G}(w,\mu^{2})
$$
\begin{equation}
\times \hat{C}_{k,g}(v-w-\ln 1/a,\xi)dw, \quad k=2, L,
\end{equation}
Using the Laplace transform method, we can turn the above equation from $v$-space into $r$-space as follows:
\begin{equation}
f_{k,g}^{q\bar{q}}(r,Q^{2},m_{q}^{2})=e_{H}^{2}g(r,\mu^{2})h_{k,g}(r,\xi), \quad k=2,L, 
\end{equation}
where $ h_{k,g}(r,\xi)=\mathcal{L}[\hat{C}(v-\ln 1/a,\xi)],v,r]$ at the LO approximation have been given in Ref. \cite{321}. To obtain the heavy quarks components $F^{q\bar{q}}_{2}$ and $F^{q\bar{q}}_{L}$ of the structure functions in Laplace space at the LO and NLO approximations, the obtained gluon distribution function  in Eq. (8) are inserted into Eq. (11). But before that, $Q^{2}$ must be replaced by $\mu^{2}$.  With these descriptions, one can write the structure functions as follows:
$$
f_{k,g}^{q\bar{q}}(r,Q^{2},m_{q}^{2})=e_{H}^{2}\Bigg[j_{gf}(r,\mu^{2})f(r,Q_{0}^{2})
$$
\begin{equation}
+ j_{gg}(r,\mu^{2})g(r,Q_{0}^{2})\Bigg], \quad k=2, L,
\end{equation}
where
$$
j_{gf}(r,\mu^{2})= h_{k,g}(r,\xi)k_{fg}(r,\tau(\mu^{2},Q_{0}^{2}))/a^{r},
$$
\begin{equation}
j_{gg}(r,\mu^{2})= h_{k,g}(r,\xi)k_{gg}(r,\tau(\mu^{2},Q_{0}^{2}))/a^{r}.
\end{equation}
Finally, using the Laplace inverse transform, we can obtain these structure functions in the usual space $x$ as follows;
$$
F_{k,g}^{q\bar{q}}(x,Q^{2},m_{q}^{2})=e_{H}^{2}\Bigg[J_{gf}(x,\mu^{2})\otimes F_{s}(x,Q_{0}^{2})
$$
\begin{equation}
+ J_{gg}(x,\mu^{2})\otimes G(x,Q_{0}^{2})\Bigg], \quad k=2, L
\end{equation}
where $J_{gf}(x,\mu^{2})=\mathcal{L}^{-1}[j_{gf}(r,\mu^{2}),r,v]\vert_{v=\ln(1/x)}$ and $J_{gg}(x,\mu^{2})=\mathcal{L}^{-1}[j_{gg}(r,\mu^{2}),r,v]\vert_{v=\ln(1/x)}$. It should be noted that, to obtain the heavy quarks structure functions in Eq. (14), it is requires only a knowledge of the singlet $F_{s}(x)$ and gluon $G(x)$  distribution functions at the starting value $Q^{2}_{0}$.
\section{numerical Results }
Now, we present our numerical results of the production cross section of heavy quarks in the $e^{-}p$ interaction at the LO and NLO approximations obtained  by using Eq. (1) and the DGLAP evolution equations.  In order to present more detailed discussions on our findings, the numerical results for the structure functions and the production cross section of heavy quarks are compared with HREA data \cite{8,9,13,15,17,42,421,422} and with the results from the MSTW2008 \cite{48}, MSRT03 \cite{49}  and H1PDF  \cite{61}. To extract numerical results, we use the published CT18 \cite{63} initial starting functions $F_{s}(x)$ and $G(x)$. It should be said that, we consider the uncertainties due to the running charm, beauty and top (t) quark masses $m_{c} = 1.29^{+0.077}_{-0.053}$ $GeV$, $m_{b} =4.049^{+0.138}_{-0.118}$ $GeV$  \cite{10} and $m_{t} =173.5^{+3.9}_{-3.8}$ $GeV$ \cite{60}  where the uncertainties are obtained through adding the experimental fit, model and parameterization uncertainties in quadrature.

Figures (1) and (2) indicate the numerical results of the c quark structure functions ($F^{c\bar{c}}_{2}(x,Q^{2})$ and $F^{c\bar{c}}_{L}(x,Q^{2})$) at $Q^{2}=11, 60, 130$ and $500$ $GeV^{2}$ and the b quark structure functions ($F^{b\bar{b}}_{2}(x,Q^{2})$ and $F^{b\bar{b}}_{L}(x,Q^{2})$) at $Q^{2}=12, 60, 200$ and $650$ $GeV^{2}$. These results are presented at the LO and NLO approximations and compared with those presented by ZEUS \cite{42,421,422} and H1 \cite{13,15,154} colliders and with  the results from the MSTW2008 predictions \cite{48}. In these figures, since nowhere is presented data on the longitudinal structure function of heavy quarks, we only compare $F^{c\bar{c}}_{2}(x,Q^{2})$ and $F^{b\bar{b}}_{2}(x,Q^{2})$  with those presented by ZEUS and H1 colliders . As can be seen, our numerical results at the NLO approximation are closer to the experimental data than the results at the LO approximation. 

In figure (3) as a comparison, we show our numerical results of the c quark reduced cross section and ZEUS \cite{9} data  as a function of $x$. The numerical results of this cross section are showed at the LO and NLO approximation at $Q^{2}=6.5, 12, 30, 80,160$ and $600$ $GeV^{2}$.  It can be concluded that in the low-energy scale, our results are very close to ZEUS data and show that the presented control coefficients are suitable. We also compare the c quark reduced cross section at the NLO approximation in $Q^{2}=5, 12, 60, 200, 650$ and $2000$ $GeV^{2}$ with H1 \cite{8} data and with the results from the MSTW2008  predictions \cite{48} in figure (4). 

Figure (5) indicates our numerical results of the b quark reduced cross section at the LO and NLO approximations in $Q^{2}=6.5, 12,30,80, 160$ and $600$ $GeV^{2}$ compared with ZEUS \cite{9} data. Moreover,  in figure (6), we compare the b quark reduced cross section at the NLO approximation  with H1 \cite{8} data and with the results from the MSTW2008  predictions \cite{48}. 

In figure (7), we present the numerical results of the t quark reduced cross section at the LO and NLO approximations at $Q^{2}=6.5, 12,30,80, 160$ and $600$ $GeV^{2}$. Here, we must state that the t quark longitudinal structure function at the LO and NLO approximations is very small relative to the structure function $F^{t\bar{t}}_{2}(x,Q^{2})$ and the value of this function at the specified energies does not have a significant effect on the t quark reduced cross section.

In figure (8), it is presented a comparison between the reduced cross section of heavy quarks at $Q^{2}=1000,5000$ and $10000$ $GeV^{2}$. In this figure at $Q^{2}=1000GeV^{2}$ and minimum $x$, the ratios of  $\sigma^{b\bar{b}}_{red}/\sigma^{c\bar{c}}_{red}$ and $\sigma^{t\bar{t}}_{red}/\sigma^{c\bar{c}}_{red}$ are approximately $0.11$ and $0.0002$, respectively and at $Q^{2}=10000GeV^{2}$ and minimum $x$, these ratios are approximately $0.18$ and $0.0035$, respectively. These results show that with the increase of energy, the production cross section of t quark grows more than the production cross section of b quark.

In order to assess the significance of the theoretical uncertainty at the LO and NLO approximations, we indicate the $Q^{2}$ dependence of the single differential cross section of  heavy quarks $d\sigma^{qq}/dQ^{2}$ at the LO and NLO approximations in figure (9). In this figure, the differential cross section of c, b and t quarks are presented at the center of mass energies of $\sqrt{s}=319GeV$, $\sqrt{s}=1.3TeV$ and $\sqrt{s}=3.5TeV$ in $e^{-}p$ interaction at $0.1<y<0.7$. We also show the single differential cross section of the c and b quarks as a function of $Q^{2}$ at the CME of $\sqrt{s}=319GeV$ and  $0.02<y<0.7$ at the NLO approximation in figure (10). The data have been
given together with their statistical and systematic uncertainties (not including the error on the integrated luminosity). Moreover,  the single differential cross sections of c and  b quarks as a function of $x$ at the CME of $\sqrt{s}=319GeV$ and  $0.02<y<0.7$ at the NLO approximation are presented in figure (11).

In Table (2), the integrated cross sections are compared with H1 \cite{31} and ZEUS \cite{17} data  and with the predictions from NLO QCD. The integrated cross sections for c and b quarks have been respectively presented  $ 373 \pm 39 \pm 47$ $pb$ and  $55.4\pm8.7\pm12.0$ $pb$ by the H1 vertex detector for values of photon virtuality $Q^{2}> 150 GeV^{2}$ and of inelasticity $0.1 <y< 0.7$. Our numerical result of this cross section at the LO and NLO approximations for the c quark at the $e^{-}p$  CME of  $\sqrt{s}=319$ $GeV$ and at $Q^{2}> 150$ $GeV^{2}$ and  inelasticity $0.1 <y< 0.7$ are $312\pm7$ $pb$ and $331\pm3$ $pb$, respectively and for the b quark they are $30.9\pm1.1$ $pb$ and $35.7\pm1.0$ $pb$.  These results are presented in Table (2) and compared with the VFNS predictions from MRST03 \cite{49} and the H1PDF 2000 ﬁt \cite{61}. In addition to these results, we show the integrated cross sections for c and b quarks at the $e^{-}p$  CME of  $\sqrt{s}=1.3 TeV$ and $\sqrt{s}=3.5 TeV$. It should be noted that the integrated cross sections for c and b quarks at $\sqrt{s}=319GeV$ at the NLO approximation are larger than those at the LO approximation but at $\sqrt{s}=1.3 TeV$ and $\sqrt{s}=3.5 TeV$ the results are inverse.
Furthermore, we obtain and present the integrated cross sections for t quark at $\sqrt{s}=1.3 TeV$ and $\sqrt{s}=3.5 TeV$ in Table (2) as a prediction from our calculations. This cross section at $\sqrt{s}=319GeV$  for values of photon virtuality $Q^{2}> 150 GeV^{2}$ and of inelasticity $0.1 <y< 0.7$ is zero.

All of the results clearly show that the extraction procedure provides correct behaviors of the structure functions and the production cross section of the heavy quraks  at the LO and NLO approximations. Moreover, it should be noted that the NLO corrections are small  for values of high $x$ ,  but at low $x$ region  these corrections have many effects on the results especially at low $Q^{2}$. Furthermore,  they often allow one to reduce the uncertainties of the predicted results, as one can see by comparing the bands in almost all of the plots presented in the figures.
It should be emphasized that here we have only obtained the cross section of heavy quark production in the the  photon fragmentation region, because at high energies like $\sqrt{s}=3.5$ $TeV$, the contribution of heavy quark pair production via the gluon-gluon splitting must also be considered. 

\begin{figure}[h]
\begin{center}
\includegraphics[width=.45\textwidth]{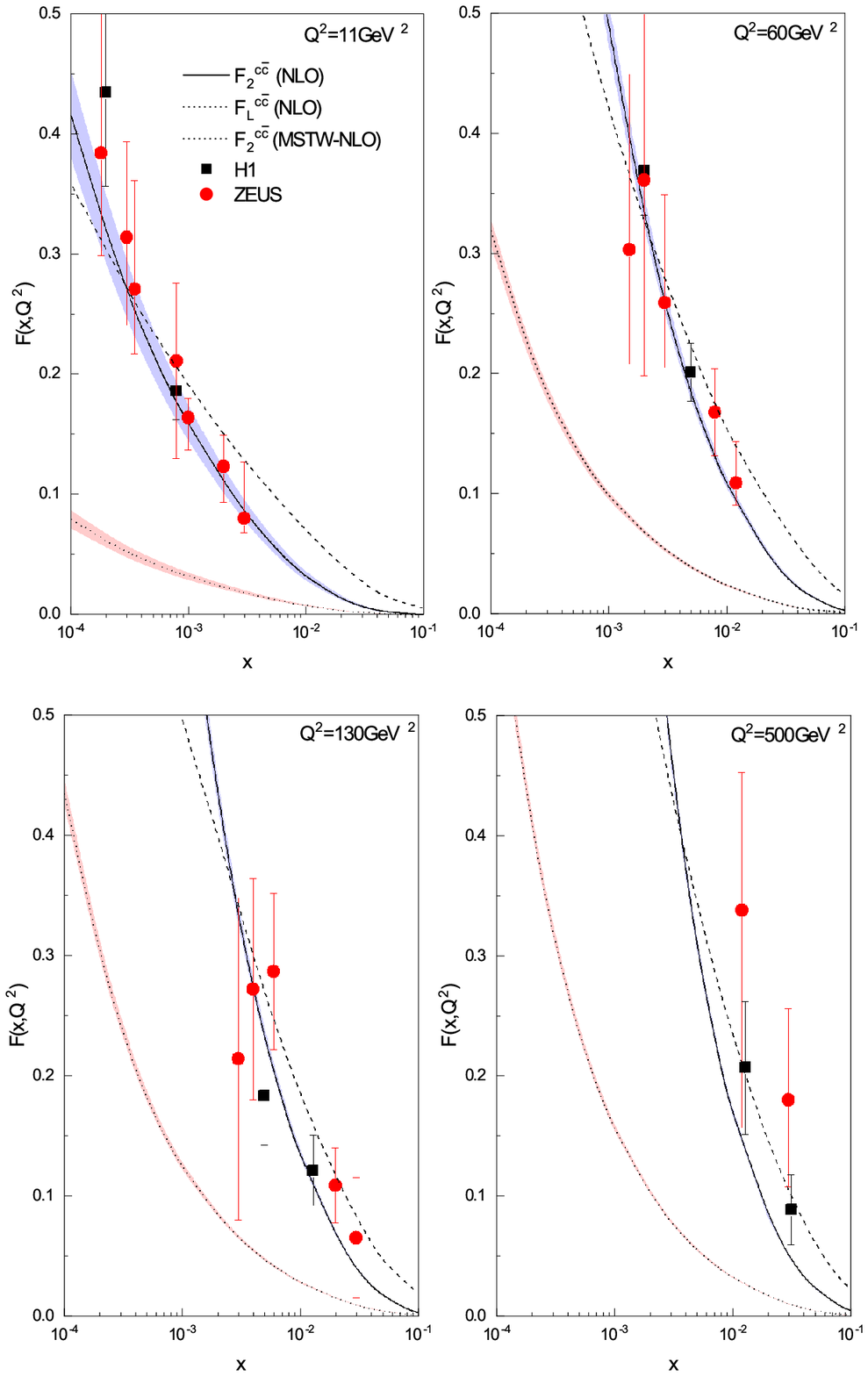}
\end{center}
\begin{center}
\selectfont{\label*{Figure (1): The charm quark structure functions $F^{c\bar{c}}_{2(L)}$ compared with data from H1 \cite{13,15,154}, ZEUS \cite{42,421,422}, and MSTW2008 \cite{48} at the NLO approximation.}}
\end{center}
\end{figure}
\begin{figure}[h]
\begin{center}
\includegraphics[width=.45\textwidth]{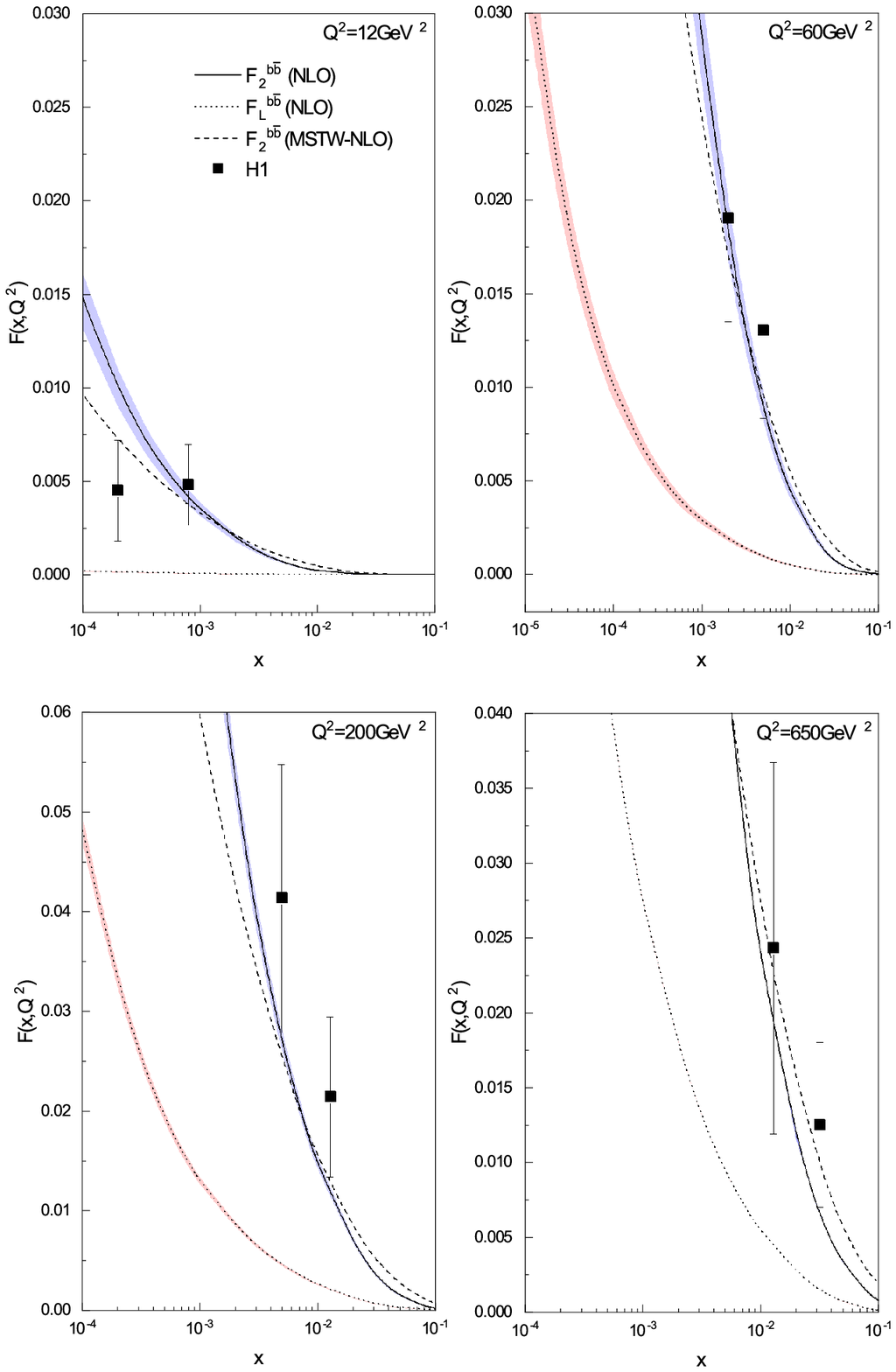}
\end{center}
\begin{center}
\selectfont{\label*{Figure (2): The beauty quark structure functions $F^{b\bar{b}}_{2(L)}$ compared with data from H1 \cite{13} and MSTW2008 \cite{48} at the NLO approximation.}}
\end{center}
\end{figure}
\begin{figure}[h]
\begin{center}
\includegraphics[width=.5\textwidth]{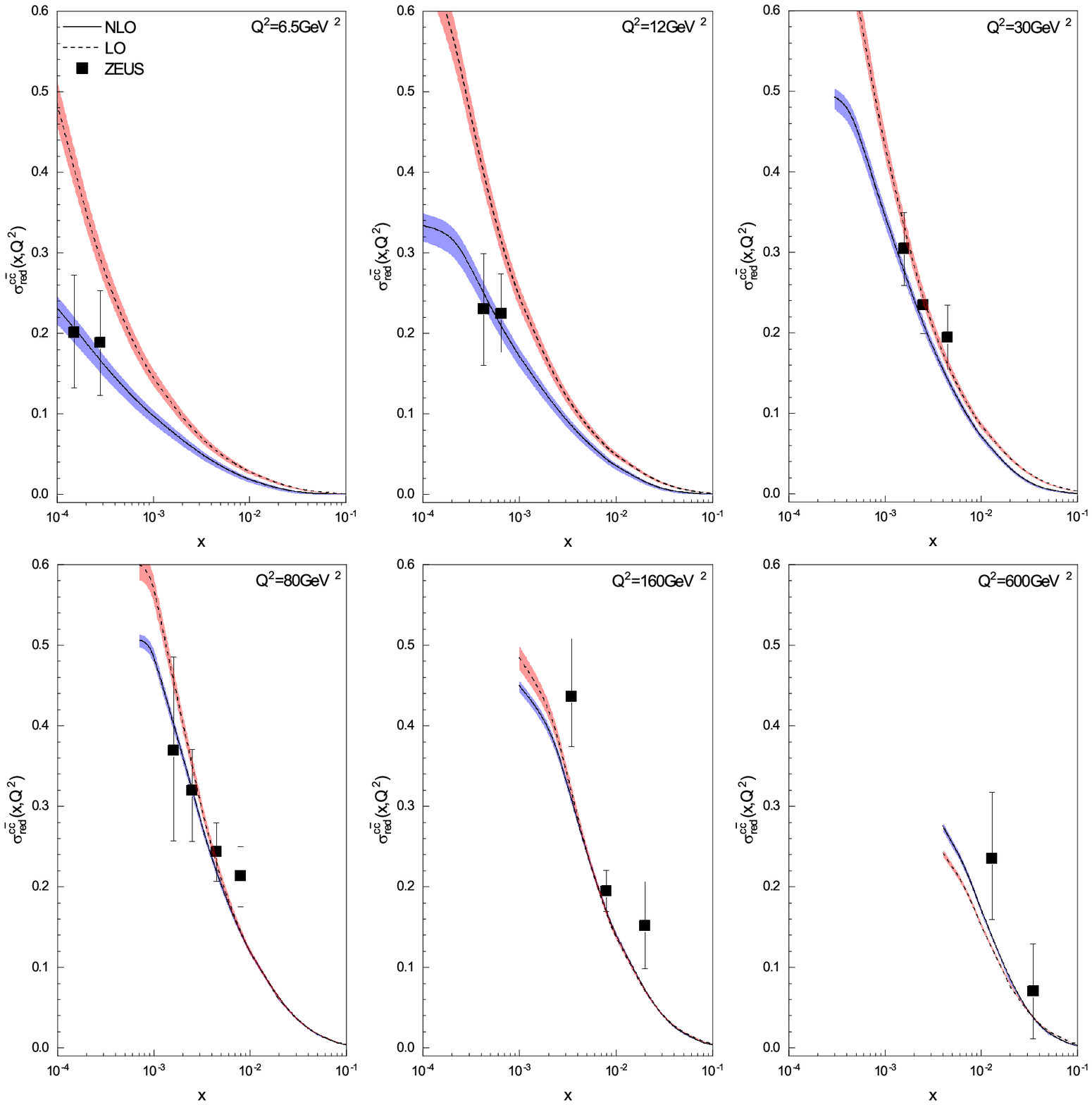}
\end{center}
\begin{center}
\selectfont{\label*{Figure (3): The reduced charm quark cross section as a function of $x$ for six different
values of $Q^{2}$ at the LO and NLO approximations compared with the ZEUS  data \cite{9}. The error bars represent the statistical, systematic (not including the error on the integrated luminosity) and extrapolation uncertainties added in quadrature. The shaded areas are the uncertainties due to the running quark mass.}}
\end{center}
\end{figure}

\begin{figure}[h]
\begin{center}
\includegraphics[width=.5\textwidth]{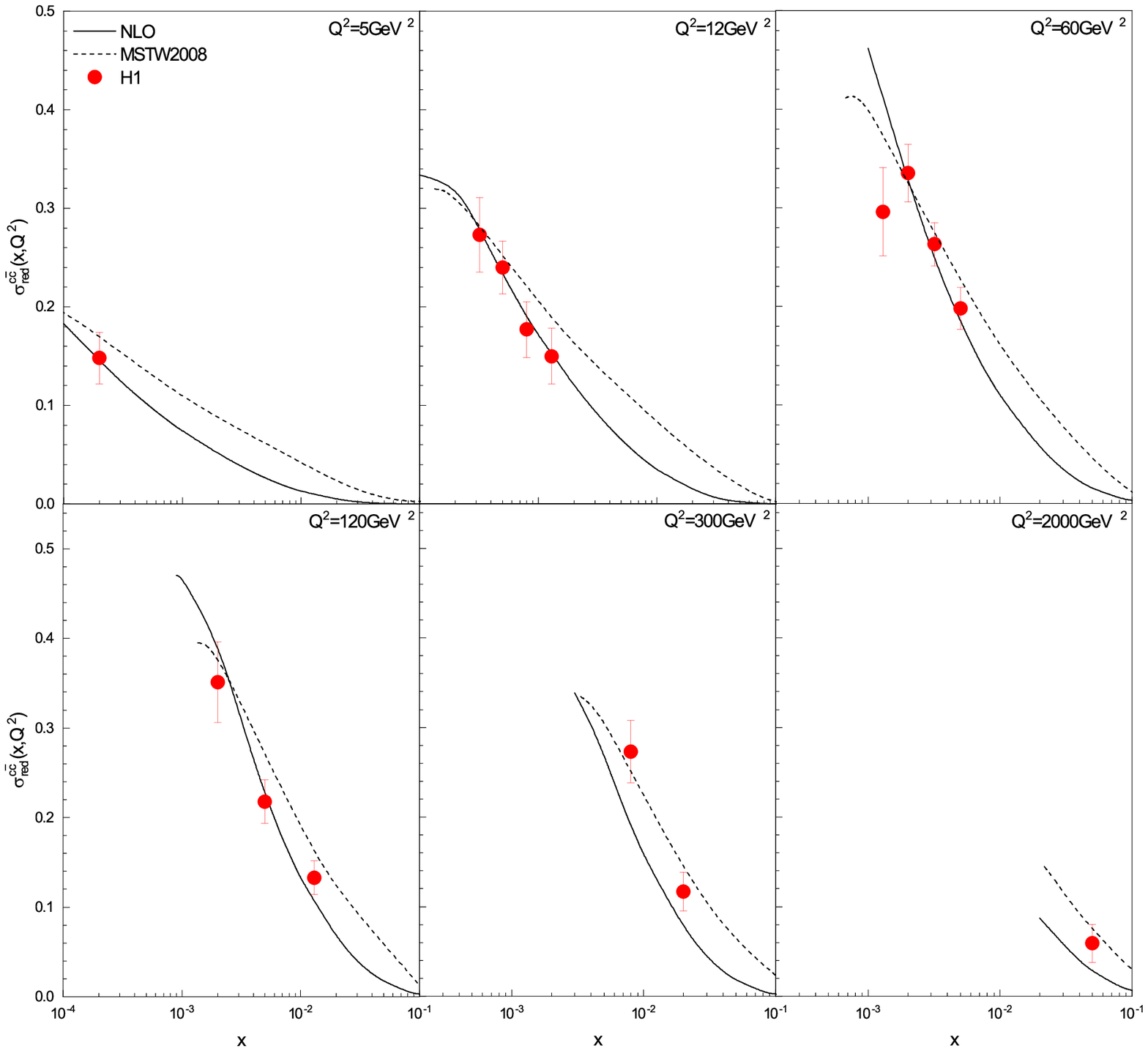}
\end{center}
\begin{center}
\selectfont{\label*{Figure (4): The reduced charm quark cross section as a function of $x$  at the  NLO approximation compared with the H1  data \cite{8} and the results from the MSTW2008  predictions \cite{48}.}}
\end{center}
\end{figure}
\begin{figure}[h]
\begin{center}
\includegraphics[width=.5\textwidth]{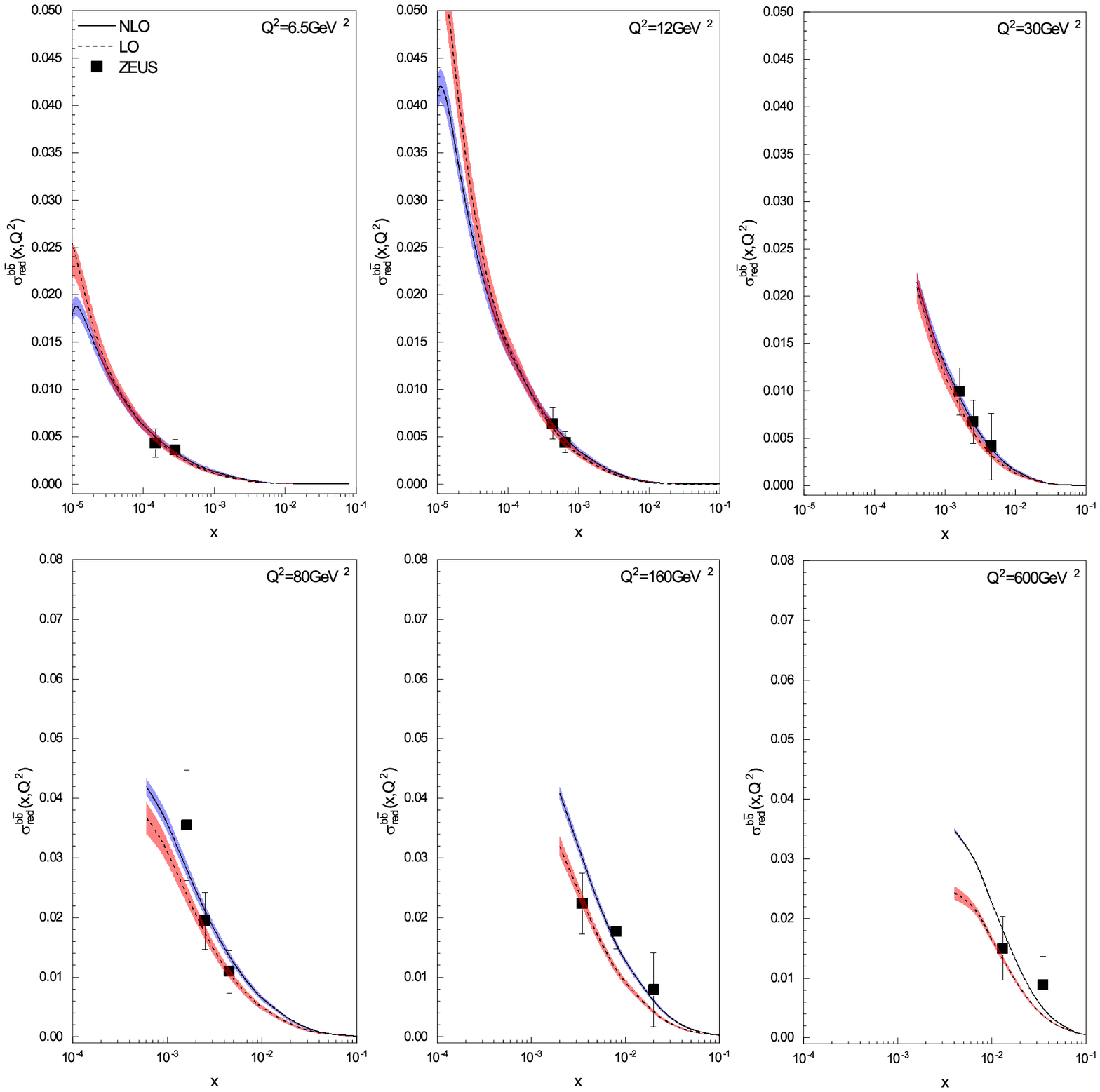}
\end{center}
\begin{center}
\selectfont{\label*{Figure (5): The reduced beauty quark cross section. For more details, see the caption of figure (3).}}
\end{center}
\end{figure}

\begin{figure}[h]
\begin{center}
\includegraphics[width=.5\textwidth]{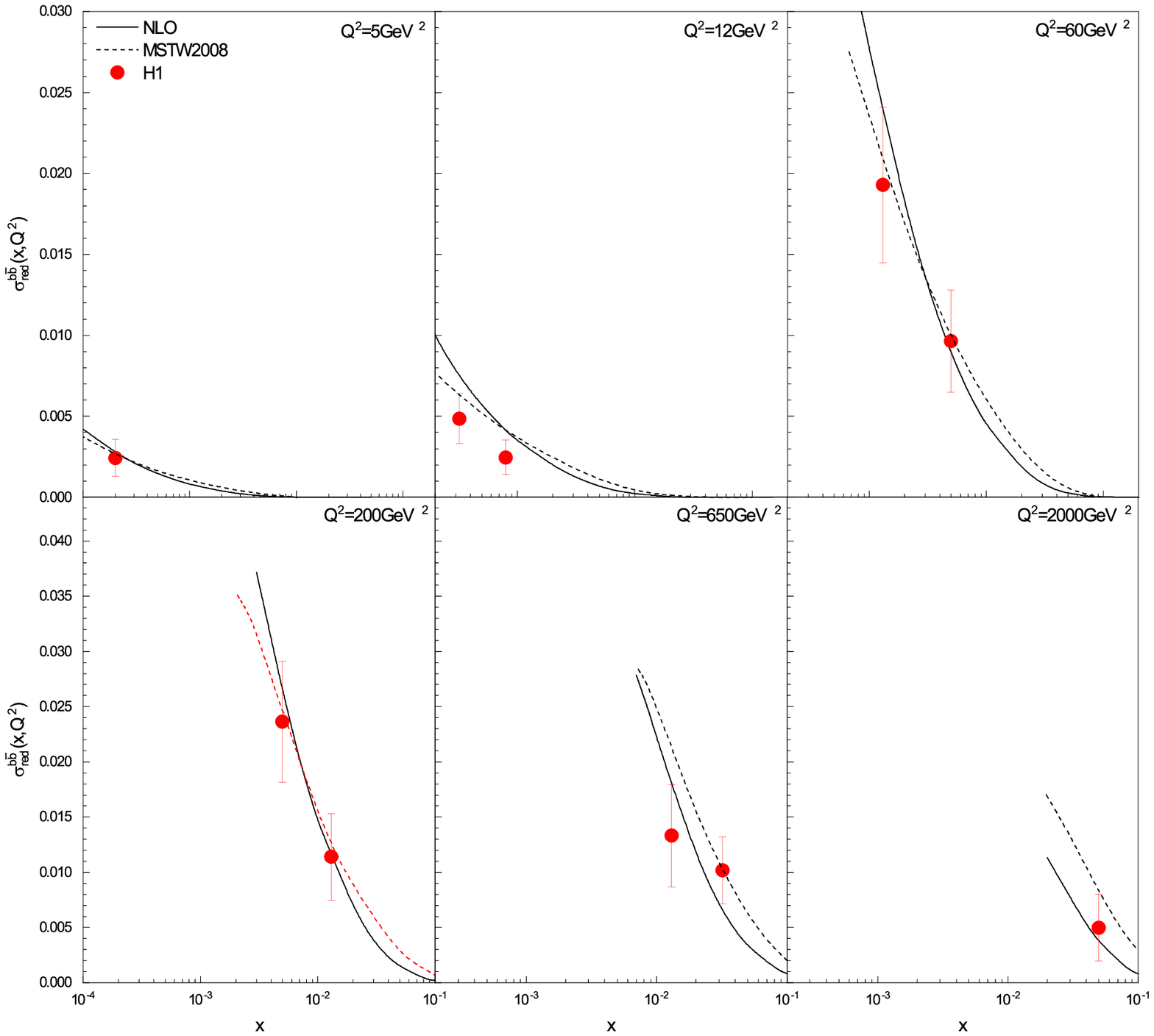}
\end{center}
\begin{center}
\selectfont{\label*{Figure (6):  The reduced beauty quark cross section as a function of $x$  at the  NLO approximation compared with the H1 data \cite{8} and the results from the MSTW2008 predictions \cite{48}.}}
\end{center}
\end{figure}
\begin{figure}[h]
\begin{center}
\includegraphics[width=.65\textwidth]{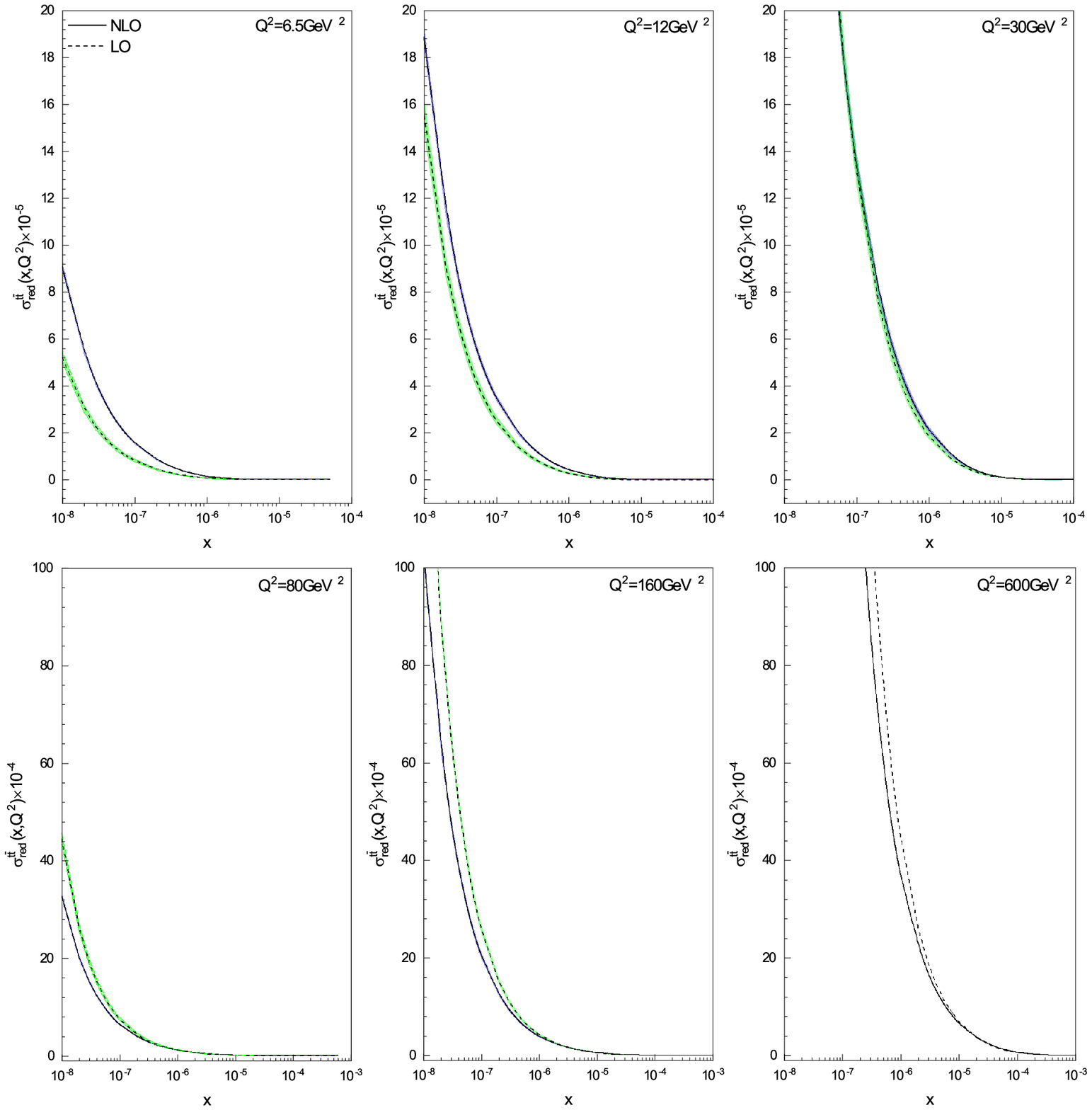}
\end{center}
\begin{center}
\selectfont{\label*{Figure (7): The reduced top quark cross section as a function of $x$ for six different
values of $Q^{2}$ at the LO and NLO approximations.}}
\end{center}
\end{figure}
\begin{figure}[h]
\begin{center}
\includegraphics[width=.5\textwidth]{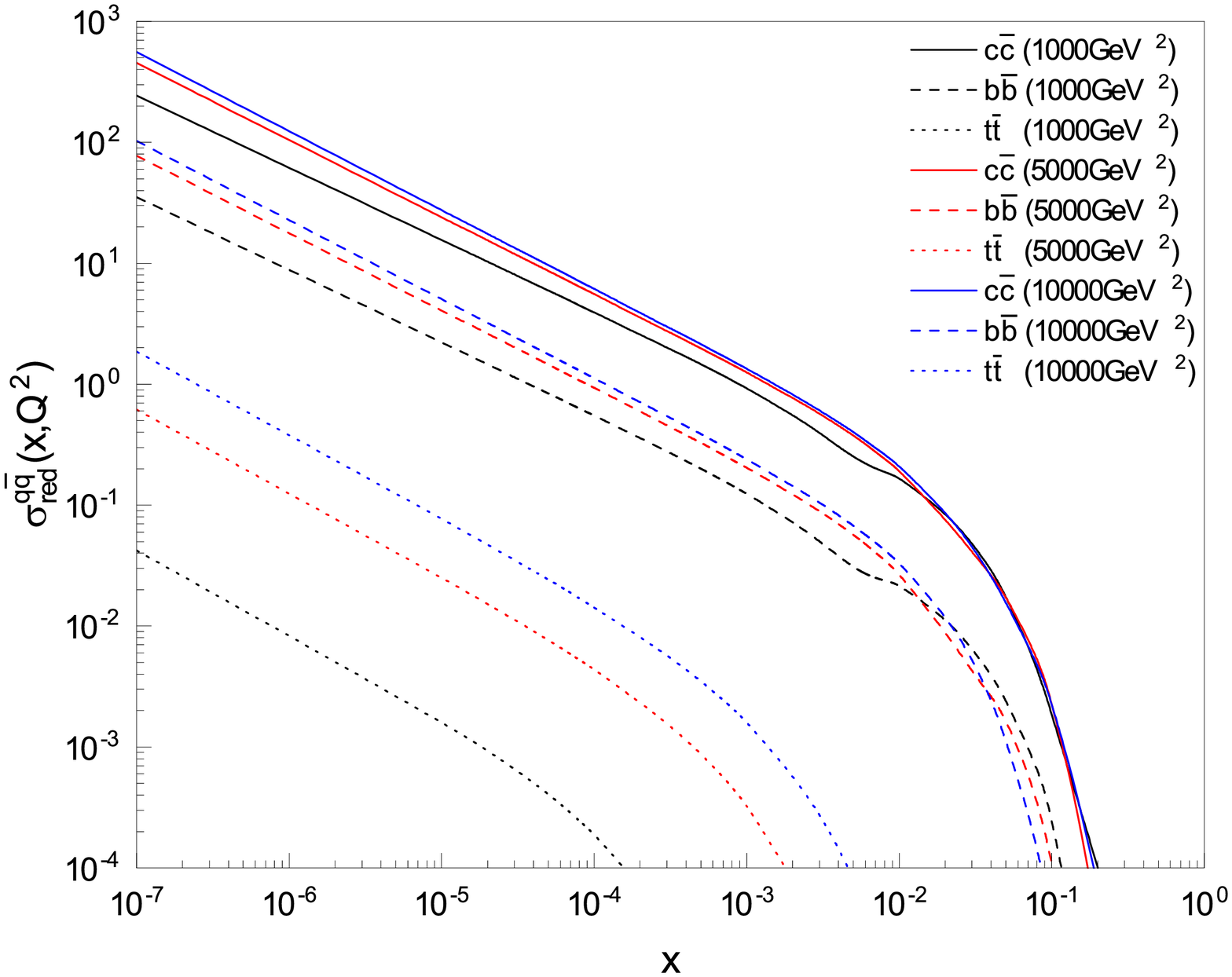}
\end{center}
\begin{center}
\selectfont{\label*{Figure (8): A comparison between the reduced cross section of heavy quarks at large values of $Q^{2}$ at the NLO approximation. }}
\end{center}
\end{figure}

\begin{figure}[h]
\begin{center}
\includegraphics[width=.5\textwidth]{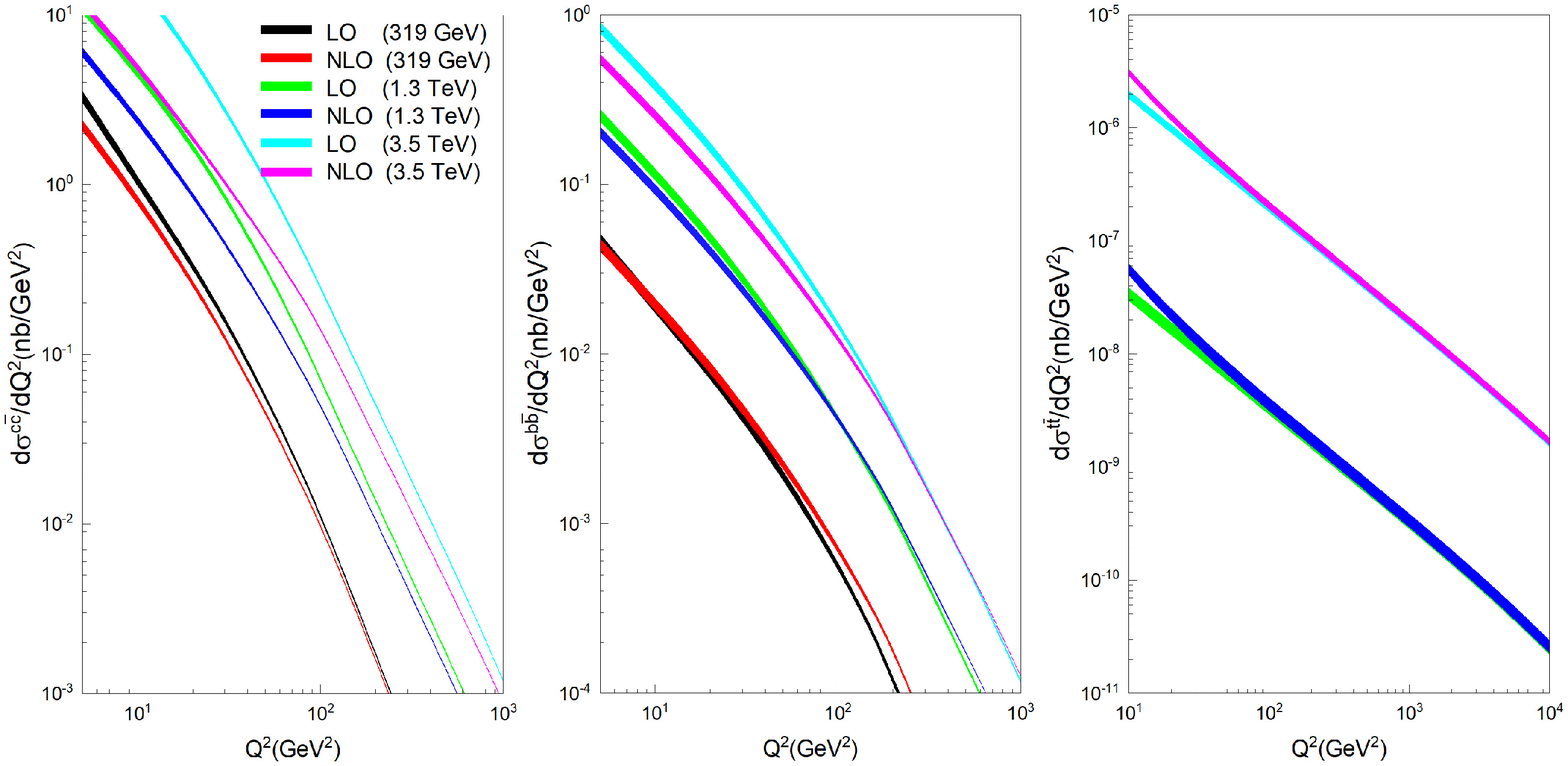}
\end{center}
\begin{center}
\selectfont{\label*{Figure (9): The results of  the  differential cross section of the c, b and t quarks as a function of $Q^{2}$  at the center-of-mass energies of  $\sqrt{s}=319GeV$, $\sqrt{s}=1.3TeV$ and  $\sqrt{s}=3.5TeV$ at the LO and NLO approximations.}}
\end{center}
\end{figure}
\begin{figure}[h]
\begin{center}
\includegraphics[width=.5\textwidth]{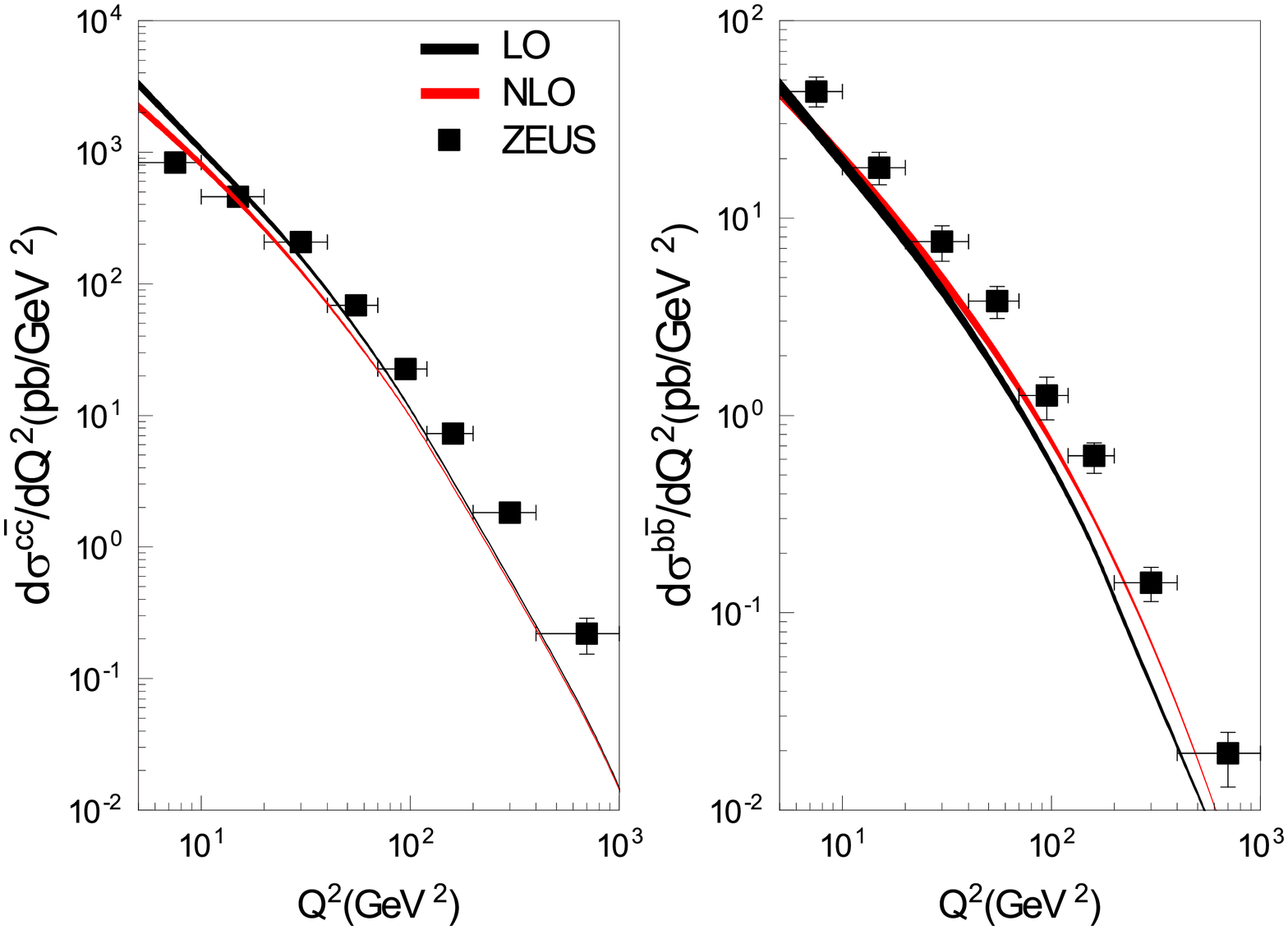}
\end{center}
\begin{center}
\selectfont{\label*{Figure (10):  The results of  the  differential cross section of the c and b quarks as a function of $Q^{2}$ at the center-of-mass energy of  $\sqrt{s}=319GeV$ at the LO and NLO approximations compared with ZEUS data \cite{9}.  }}
\end{center}
\end{figure}
\begin{figure}[h]
\begin{center}
\includegraphics[width=.5\textwidth]{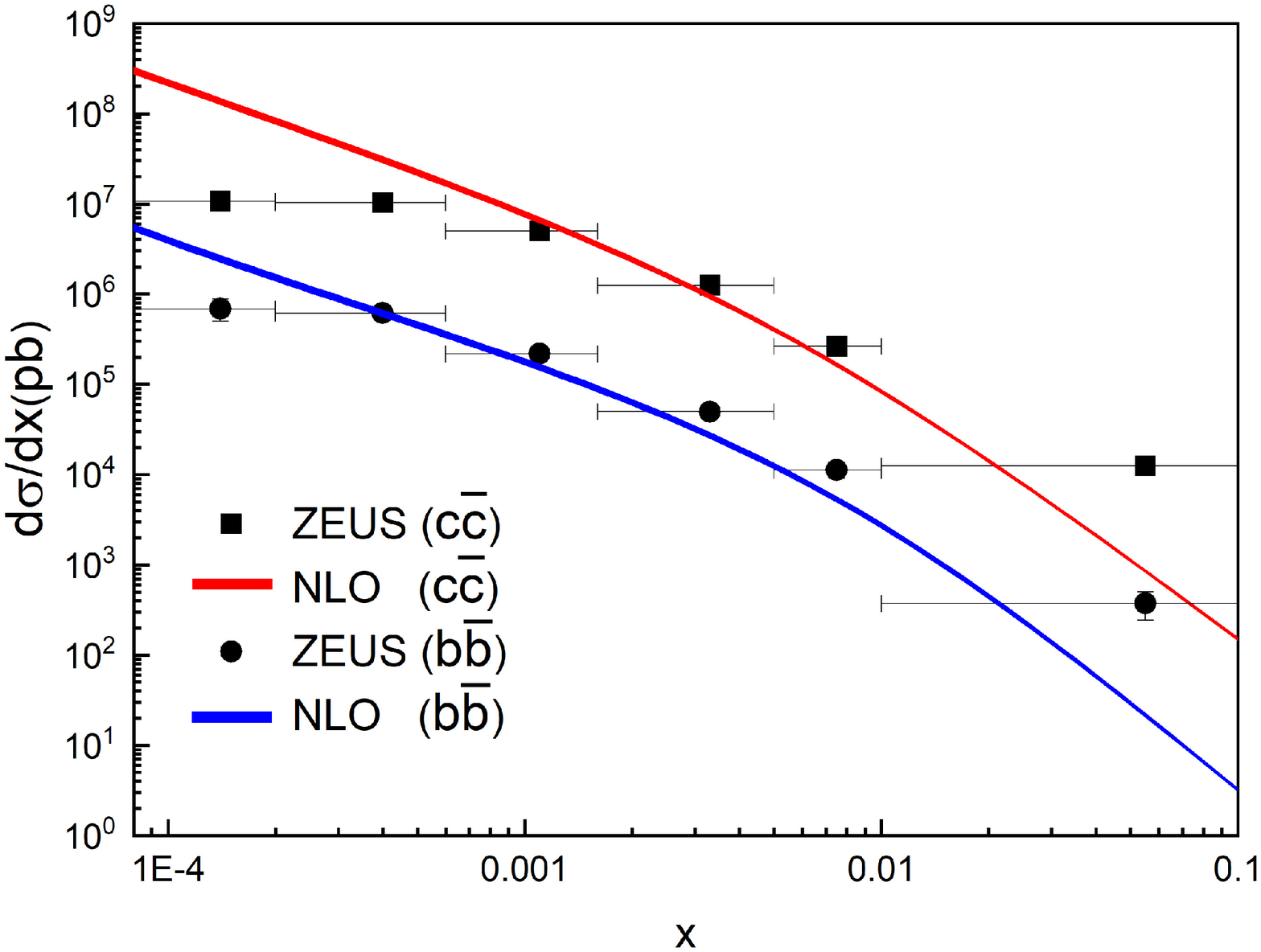}
\end{center}
\begin{center}
\selectfont{\label*{Figure (11):  The results of  the  differential cross section of the c and b quarks as a function of $x$ at the center of mass energy of  $\sqrt{s}=319GeV$ at the NLO approximation compared with ZEUS data \cite{9}. }}
\end{center}
\end{figure}
\begin{table}[h]
\begingroup
\fontsize{10pt}{12pt}\selectfont{
\label*{ Table (2): The integrated cross sections of the c, b and t quarks which are integrated over the range $Q^{2}> 150$ $GeV^{2}$ and $0.1<y< 0.7$. Our numerical results for the $c$ and $b$ quark cross sections are compared with the H1 \cite{13} and ZEUS \cite{17} data and with the results from H1PDF \cite{61} and MSRT03 \cite{49}.  }}
\newline
\endgroup
\begingroup
\fontsize{8pt}{12pt}\selectfont{
\begin{tabular} {lllll}
\toprule[1pt]
&\normal{\head{$ \sqrt{s}=319GeV$}}&
\normal{\head{$ \sqrt{s}=1.3TeV$}}\  & \normal{\head{$ \sqrt{s}=3.5TeV$}}  \ \\ \hline\\
$\sigma^{c\bar{c}}$ $(pb)$& & & \\
LO   & $312\pm7$ &  $2305\pm45$  &  $11774\pm217$ \\ 
NLO  & $331\pm3$ &  $2044\pm28$  &  $7238\pm115$ \\ 
H1  & $ 373\pm39\pm47$ &  $---$  &  $---$ \\
H1PDF  & $ 455$ &  $---$  &  $---$ \\
ZEUS & $419$ &  $---$  &  $---$ \\
MRST03 & $426$ &  $---$  &  $---$ \\ \\
$\sigma^{b\bar{b}}$ $(pb)$& & & \\
LO    & $30.9\pm1.1$ &  $251.8\pm8.3$  &  $931.7\pm28.4$ \\ 
NLO  & $35.7\pm1.0$ &  $283.7\pm6.1$  &  $917.1\pm21.1$ \\ 
H1  & $ 55.4\pm8.7\pm12.0$ &  $---$  &  $---$ \\
H1PDF  & $ 52$ &  $---$  &  $---$ \\
ZEUS & $37$ &  $---$  &  $---$ \\
MRST03 & $47$ &  $---$  &  $---$ \\ \\

$\sigma^{t\bar{t}}$ $(fb)$& & & \\
LO   & $---$ &  $1.33\pm0.18 $  &  $79.1\pm6.2$\\
NLO  & $---$ &  $1.39\pm0.17$  &  $81.2\pm6.5$ \\

\bottomrule[1pt]
\end{tabular} }
\endgroup
\end{table}

\section{Conclusion}

In conclusion, we have presented the production cross section of  heavy quarks ($\sigma^{c\bar{c}}$, $\sigma^{b\bar{b}}$ and $\sigma^{t\bar{t}}$) and the single differential cross sections ($d\sigma^{qq}/dQ^{2}$ and $d\sigma^{qq}/dx$) of them by utilizing the heavy quarks structure functions $F_{2}^{q\bar{q}}$ and $F_{L}^{q\bar{q}}$ obtained by Dokshitzer-Gribov-Lipatov-Altarelli-Parisi evolution equations and  a suitable fit for the  heavy quarks coefficient functions at the NLO approximation. Indeed,  we have shown that the Laplace transform method is the suitable and alternative scheme to solve the DGLAP evolution equations and Eq. (2). It should be noted that, the obtained equations are general and require only a knowledge of the parton distribution functions $F_{s}(x)$, $G(x)$ at the starting value $Q^{2}_{0}$. The comparisons have shown that our numerical results of the charm and beauty production cross section are in agreement with the H1 and ZEUS data well within errors. Also, in this paper, we have compared the production cross sections at the center-of-mass energy of $\sqrt{s}=319GeV$ and at $0.1<y< 0.7$  with the experimented results by H1 PDF 2000 and MSRT03 and MSTW2008. Also, we have obtained the production cross sections of heavy quarks at the-center-of mass energies of  $\sqrt{s}=1.3TeV$ and $\sqrt{s}=3.5TeV$ and considered the uncertainties due to the running charm, beauty and top quark masses. In addition, we have presented the production cross section of the quark top  at center-of-mass energies of  $\sqrt{s}=1.3TeV$ and $\sqrt{s}=3.5TeV$.


\end{document}